\documentstyle[12pt]{article}

\begin{document}

\begin{center}

{\Large \bf Effect of noise on coupled chaotic systems}

\end{center}

\vspace{5pt}

\begin{center}

Manojit Roy \footnote{email: mroy@physics.unipune.ernet.in} and
R. E. Amritkar \footnote{email: rea@physics.unipune.ernet.in}

\vskip 0.3in

Department of Physics \\
University of Pune \\
Pune--411007, INDIA.

\end{center}

\vspace{5pt}

\begin{abstract}

Effect of noise in inducing order on various chaotically evolving
systems is reviewed,
with special emphasis on systems consisting of coupled chaotic elements.
In many situations it is observed that the uncoupled elements when driven
by identical noise, show synchronization phenomena where
chaotic trajectories exponentially 
converge towards a single noisy trajectory, independent of the initial
conditions. In a random neural network, with infinite range coupling, chaos
is suppressed due to noise and the system evolves towards a fixed point.
Spatiotemporal stochastic resonance phenomenon has been observed in a
square array of coupled threshold devices where a temporal characteristic of
the system resonates at a given noise strength. In a chaotically evolving
coupled map lattice
with logistic map as local dynamics and driven by identical noise at
each site, we report that the number of {\it structures} (a structure is a 
group of neighbouring lattice sites for whom values of the variable follow
certain predefined pattern) follow a power--law decay with
the length of the structure.
An interesting phenomenon, which we call {\it stochastic coherence},
is also reported in which the abundance and lifetimes of these
structures show characteristic peaks
at some intermediate noise strength. 

\end{abstract}

\newpage

\section{Introduction}

Chaos in natural and human--made systems is a well established
fact by now. Systems in diverse
disciplines such as population biology \cite{RM}, physiology \cite{MG},
hydrodynamics \cite{LM}, chemical reactions \cite{SS}, plasma \cite{CW},
lasers \cite{TN}, electronics \cite{PL}, computing networks \cite{KHH},
economic theory \cite{HHY}, social theory etc. have been observed to exhibit
rich and unpredictable behaviour of chaos.
This behaviour has been identified with the inherent nonlinear nature
of the systems rather than external influences.

On the other hand, the very same nonlinearity has been seen to
give rise to ordering phenomena \cite{LMT},
{\it e.g.}, regular formation of cloud 
patterns, a variety of patterns in hydrodynamic flow, oscillatory patterns
in chemical reactions, in behaviour of lasers, pulse propagation in Gunn
diode etc. These patterns may be spatial, temporal, or spatiotemporal
in nature and their understanding is of very special interest.
One particularly interesting ordering phenomenon is the generic existence of
different {\it structures} in a turbulent fluid [11--18].
They originate and degenerate
randomly in space and time. These structures appear in spite of the fact
that the fluid is undergoing a turbulent
evolution and no clear understanding of this phenomenon
is as yet achieved.

Noise has been known to play a detrimental role in
many experimental situations. This motivated researchers
to develop better techniques and methods to minimize, if not 
totally remove, the effect of noise and enhance signal--to--noise ratio
and hence system performance. Slight amount of noisy perturbation
is known to destroy delicate patterns in spatially extended systems. 

In light of these facts, interest grew when evidences to
the contrary started appearing as regards the effect of noise on the ordering
phenomena. Firstly, it has been observed that addition of noise 
of a given strength to certain nonlinear systems increases the system's
response at a particular time scale, thereby improving
the signal--to--noise
ratio; these findings opened up an entirely new field of research,
known as {\it stochastic resonance} [19--27].
Secondly, noise has been seen to influence spatial and 
spatiotemporal behaviour of some nonlinear systems in quite
counterintuitive manner. It is observed that noise can trigger,
select and sustain patterns in optical systems,
fluid dynamical systems etc. [28--32].

Considerable progress has been made to establish connection between
these pattern forming systems (such as fluid) and nonlinear dynamical
systems' theory. A natural next step is to see how does noise influence
spatiotemporal evolution of these dynamical systems, with the hope that
this kind of study may shed some light on the abovementioned
noise--induced features in physical systems. This article
presents a review of the work that has been carried out
on the effect of noise in inducing order in an otherwise
chaotically evolving system, with particular emphasis on systems consisting
of coupled chaotically evolving elements.

We have organized the article as follows: Section 2 discusses 
noise induced ordering phenomena observed in low
dimensional uncoupled systems. In Section 3, we review the work on effect
of noise on two spatially extended systems, neural network and array of
firing elements. Section 4 introduces coupled map lattice (CML), one of
the most popular models of nonlinear dynamical systems with spatial extension.
In this same section we discuss a recent work that we have carried out on the 
effect of noise on CML. Section 5 summarises and concludes
the article.

\section{Effect of noise on uncoupled systems}

In this section we investigate some examples of noise induced ordering
in uncoupled systems.

\subsection{Synchronization}

In some systems different trajectories  
get synchronized asymptotically to a single noisy trajectory
independent of the initial conditions when driven by an identical
sequence of noise above certain strength.
Here, synchronization should be understood
as exponential convergence of the average distance between any two phase
space points. This means that synchronization is essentially a 
nonchaotic phenomenon associated with negative Lyapunov exponent
\cite{AP1}, although the asymptotic trajectory can be very random.
(Lyapunov exponent $\lambda$ characterizes the rate at
which the distance between neighbouring trajectories changes.
If $\epsilon_0$ is the initial separation between two
trajectories, the separation after time $t$ can be written as
$\epsilon_t \simeq \epsilon_0 \exp(\lambda t)$.
Positive $\lambda$ implies exponential divergence of
nearby trajectories
and hence chaos.
For multidimensional systems our reference to 
Lyapunov exponent will always mean
the maximum of the Lyapunov characteristic exponents.)

In the continuum--time limit such systems
can be written as a set of uncoupled Langevin equations
\begin{eqnarray}
\dot{x}_i(t) = F(x_i) + \eta(t)\;, \label{le}
\end{eqnarray}
where $x_i$ is a dynamical variable, the index $i$ corresponds
to different trajectories $i.e.$ different initial conditions,
the (nonlinear) function
$F(x)$ governs the dynamics of the system, and $\eta(t)$ is
a delta--correlated
($\left< \eta(t) \eta(t^\prime) \right> = \left< \eta^2(t) \right>
\delta(t - t^\prime)$)
noisy driving force imparted at a regular interval small
compared to all relevant macroscopic time scales of the system,
and is same for all $i$.
Eqs.~(\ref{le}) have the synchronization solution
$x(t) = x_i(t)$.
Therefore the question essentially is whether this solution is
stable or not for a given range of system parameters. In other
words, if an appropriately constructed
$\lambda < 0$, there is an
exponential convergence of trajectories and hence synchronization,
and if $\lambda > 0$, the evolution is chaotic.

Let us now take up individual cases.

\subsubsection{Particle in newtonian potential}

Fahy and Hamann (FH) \cite{FH} have
studied a Newtonian particle moving
in a smooth multiminima potential $V(x)$ without friction,
subjected to the condition
that at regular time intervals $\tau$, it is stopped and all its 
velocity components are reset to random values chosen from a Gaussian
distribution. They observed that when an ensemble of such particles
with different initial conditions is driven by an identical
sequence of random forces, their trajectories asymptotically get
synchronized to a single 
noisy trajectory provided that $\tau$ is less than certain
critical value $\tau_c$.
On the basis
of this observation they concluded that for $\tau < \tau_c$ the final
trajectory of the particles is independent of the initial conditions
to any required level of accuracy (the accuracy aspect will have
importance later in our discussion); it depends only on the choice
of velocities.
Thus the trajectory, albeit noisy, is not chaotic for $\tau < \tau_c$
because of exponential convergence, and an appropriately defined
lyapunov exponent $\lambda$ is negative. They have shown
that for any one--dimensional potential $V(x)$ confining
the particles to a finite region and for short enough $\tau$, the average rate
of contraction $\gamma$ of
the distance between two particles initially close together is given
by $\gamma = 
\tau \beta \left< (\partial V/\partial x)^2 \right> /2m
 + O(\tau^2)$, where $\beta = 1/k_B T$, and $m$ is the mass of the particle
(angular brackets denote average with respect to the Boltzmann
distribution).
The above mentioned synchronization phenomenon is a stronger observation
than the well--known statistical phenomenon in which the 
asymptotic distribution of Brownian trajectories is found to be
proportional to the Boltzmann factor ${\rm exp}[- \beta V(x)]$, independent
of its initial distribution.
FH have also conjectured that this synchronization
feature may be generic to all the bounded systems.

\subsubsection{Noisy chaotic systems}

Maritan and Banavar (MB) \cite{MB1} have considered the
effect of noise on the following two chaotic systems. First,
they have taken a noisy logistic map with the evolution law
\begin{eqnarray}
x_{t+1} = 4x_t(1 - x_t) + \eta_t\;, \label{lm}
\end{eqnarray}
where $\eta_t$ is a uniform--deviate random number chosen
from an interval $[-W,$ $+W]$ with the constraint that $0 < x_{t+1} < 1$.
Eq.~(\ref{lm}) may be considered as an example of the discrete--time
version of Eq.~(\ref{le}) with $\Delta t = 1$.
They found a critical $W_c (\approx 0.5)$ such that when a pair of
randomly chosen initial conditions is driven by an identical sequence
of $\eta_t$ with $W > W_c$, their asymptotic trajectories become
synchronized (within a given accuracy) to a single random
trajectory independent of the initial conditions. They have computed the
mean squared separation $\bar{d^2}$ between the two identically driven
trajectories and observed that  
$\bar{d^2}$ falls
off for $W > W_c$.
They offered an intuitive explanation saying that the convergence can
occur for logistic map due to the contraction of $d$ whenever
sum of the pair of numbers comes close to unity, because the distance
evolves as
\begin{eqnarray}
|x_{t+1}(1) - x_{t+1}(2)| =
			4|x_t(1) - x_t(2)|[1 - \{x_t(1) + x_t(2)\}]\;.
							\nonumber
\end{eqnarray}
Since peaks in the invariant density for the logistic map are close to
0 and 1, $W \geq 1/2$ can bring both $x(1)$ and $x(2)$ near 1/2.

The second system considered by MB is the Lorenz system described by
\begin{eqnarray}
dx/dt = \sigma(x - y)\;,\;\; dy/dt = -xz + rx - y\;,\;\; dz/dt = xy - bz\;,
							\label{ls}
\end{eqnarray}
with $\sigma = 10,\; b = 8/3,\; {\rm and}\; r = 28$. The $y$--equation is
then evolved in difference form as
\begin{eqnarray}
y(t + \Delta t) = y(t) + [-x(t)z(t) + rx(t) - y(t)]\Delta t
	        + \eta_t W_l \sqrt{\Delta t}\;, \label{ls1}
\end{eqnarray}
where $\eta_t$ is again a delta--correlated uniform--deviate random
number between $[0,1]$ (it has a nonzero mean unlike in the logistic case)
and $W_l$ is the amplitude.
Note that here addition of noise is unrestricted
unlike in the logistic case where boundedness of the phase space
constrains noise to depend on the state of the system.
For Lorenz system also a threshold value $(\approx 2/3)$ for $W_l$
was found beyond which synchronization phenomenon was observed for an
identically driven system with different initial conditions. It is
reported that synchronization does not occur if $\eta_t$ with the same
amplitude has zero mean.

MB have maintained that although the systems considered above are
exhibiting
synchronization, their strange
attractors are not replaced by topologically simple structures
like fixed point or limit cycle.

Pikovsky made a cautionary observation \cite{AP2} that for
all bounded systems there
is always a nonzero probability that any two phase space points will
come close to within some $\epsilon > 0$ with or without noise. In
other words after sufficiently long time the two systems will be
synchronized because of the finite precision (of the computer) and
finiteness of phase space.
This type of spurious synchronization will occur even for systems
with positive $\lambda$, which is to be distinguished from
the physical synchronization which is characterized by negative
$\lambda$. The former is unstable against small
perturbations because of positive $\lambda$ and so will not
be observed in real experiments, whereas the latter is quite stable.
Pikovsky found $\lambda$ for noisy logistic
system of MB to be positive and concluded that this type of
synchronization is a numerical artifact.

MB subsequently noted \cite{MB2} that whether or not 
a physical synchronization occurs, identical noise with sufficient
strength drastically enhances the probability of `close encounter'
between any two ensemble points.

There has been an attempt \cite{GC} to interpret synchronization
in terms of the inherent structural instability of the undriven system. 
These authors maintain that the strange attractor of the undriven
chaotic system gets replaced by a stable fixed point
under parametric
perturbations, resulting in synchronization of any two phase space
points and a negative $\lambda$. They showed this for all the
three systems mentioned above by treating noise as a perturbation in 
the parameter and also considered new examples. It
was remarked that this phenomenon is not generic and holds for only
those systems in which such type of structural instability occurs.

The picture that is emerging out of these findings is that while
understanding of the synchronization phenomenon may still remain
incomplete, there is no doubt about noise playing a crucial role
in drastically increasing the chance of close encounter of any two
phase space points and thereby bringing down the
lyapunov exponents of the otherwise chaotic system.

\subsection{Violation of law of large numbers}

An interesting observation was reported \cite{SS1} regarding
the chaotic evolution of an ensemble of uncoupled maps driven by
a parametric noise. The system consists of $N$ local maps
\begin{eqnarray}
x_{t+1}(i) = F(x_t(i);\;a_t(i))\;,
\end{eqnarray}
where the nonlinearity parameter $a_t(i)$ is subjected to fluctuations in
both space and time.
The quantity of interest is the mean field $h_t$, defined as
\begin{eqnarray}
h_t = \frac{1}{N} \sum_{j=1}^N F(x_t(j))\;. \nonumber
\end{eqnarray}
For the uncoupled variables $x_t(i)$ fluctuating almost independently,
$h_t$ for large $N$ is expected to obey the law of large numbers,
and hence
the mean--square deviation (MSD) (= $ \left< h_t^2 \right> -
\left< h_t \right> ^2$) should
vary as $1/N$, and converge to a fixed--point value as $N \rightarrow \infty$.
It was observed that with the logistic map in chaotic regime,
MSD falls off as $\sim 1/N$ if $a_t(i)$ fluctuates in space.
On the other hand if $a_t(i)=a_t$ is independent of $i$ then
MSD saturates with $N$ beyond a critical $N = N_c$, whose value
depends on the strength of the noise. Similar behaviour was observed
using other maps
like the circle and tent maps. The author claims that an ensemble of 
uncoupled chaotic maps with spatially uniform parametric fluctuations 
violates the law of large numbers, irrespective of the details of the map.

\section{Effect of noise on spatially extended systems}

We shall now consider systems which have
a spatial extension. Because of increased complexity owing 
to the largely enhanced phase space dimensions (one mostly talks of
infinite dimensional phase spaces) these systems show very
rich dynamical features. Spatially extended systems are modelled by a
set of variables
(whose evolution may be governed by discrete--time nonlinear
maps like the logistic
map, or continuous--time ordinary differential equations such as the 
oscillator equation) physically coupled to each other in euclidean space.

Let us consider a few cases of such systems and the influence
of noise on their evolution.

\subsection{Noise and suppression of chaos in neural network}

Molgedey, Schuchhardt and Schuster (MSS) have investigated
the effect of noise on discrete--time evolution of a random neural network
with infinite--range interactions \cite{MSS}. The model consists of
$N$ analog neurons $\{x_t(i)\}$, $i = 1, \cdots, N$, with 
$-1 \leq x(i) \leq 1$, evolving according to the law
\begin{eqnarray}
x_{t+1}(i) & = & F(h_t(i))\;, \label{nn}\\
h_t(i) & = & \sum_{j \neq i} \varepsilon_{ij} x_t(j) + \eta_t(i)\;,
							\label{nn1}
\end{eqnarray}
where the function $F(h)$ has the following properties: 1. it is
odd [$F(-h) = -F(h)$]; 2. it approaches $\pm 1$ as $h \rightarrow
\pm \infty$; 3. it increases near $h = 0$ as $[dF/dh]_{h=0} = g\;,$
where $g > 0$ is the gain parameter.
The coupling parameters $\varepsilon_{ij}$ are
delta--correlated Gaussian random variables with zero mean,
$h_t(i)$ denotes
the internal field of the neuron and $\eta_t(i)$ is the external white
noise with zero mean and variance
$ \left< \eta_t(i) \eta_\tau(j) \right>
			= \sigma^2 \delta_{ij} \delta_{t \tau}\;$.

MSS used a dynamical functional approach to reduce the dynamics of the entire
system to an equation for an effective single neuron in 
the thermodynamical limit:
\begin{eqnarray}
x_{t+1} = F(h_t)\;, \label{mnn}
\end{eqnarray}
with
\begin{eqnarray}
\left< h_t h_\tau \right> & = & 
\sigma^2 \delta_{t \tau} +
			\left< F(h_{t-1}) F(h_{\tau-1}) \right>\;,
							\label{mnn1}
\end{eqnarray}
Eq.~(\ref{mnn}) with Eq.~(\ref{mnn1}) yields the same averaged dynamical
properties as for Eq.~(\ref{nn}).
One now defines activity of the network as $K_t = \left< h_t^2 \right>$
which for small $K$ reduces to
\begin{eqnarray}
K_t = \sigma^2 + g^2K_{t-1} + O(K^2_{t-1})\;. \label{ann1}
\end{eqnarray}
In the absence of noise there is a trivial fixed point $K^* = 0$
for Eq.~(\ref{ann1}) and is stable for $g < 1$.
In the presence of noise it has only one stable fixed point in the range
$[0,1+\sigma^2]$.

To study the chaotic behaviour of the network a replica of the system with
infinitesimally separated initial conditions and with the same noise
$\eta_t^1(i) = \eta_t^2(i)$ is constructed.
Lyapunov exponent for the system is
defined as
\begin{eqnarray}
\lambda = \lim_{\tau \rightarrow \infty}
	\lim_{x_t^1 \rightarrow x_t^2} \frac{1}{2 \tau} log_2
	\frac{ \left< (x_{t+\tau}^1 - x_{t+\tau}^2)^2 \right>}
			{ \left< (x_t^1 - x_t^2)^2 \right>}\;.
							\label{mlnn}
\end{eqnarray}
Assuming equilibrium
$( \left< (h_t^1)^2 \right> = \left< (h_t^2)^2 \right>)$,
they have obtained for the noiseless case
$\lambda < 0$ for $g \le 1$
and the system goes to the trivial fixed point $K^* = 0$. For
$g > 1$, $\lambda >0$ and the system shows chaos.

They have numerically studied the effect of noise on chaotic properties of
the system using the following form for $F(h)$:
\[F(h) = \left\{ \begin{array}{ll}
		  -1\;\;\;\;\;\;& \mbox{for $h < -1/g\;,$} \\
		  gh\;\;\;\;\;\;& \mbox{for $-1/g \leq h \leq +1/g\;,$} \\
		  +1\;\;\;\;\;\;& \mbox{for $+1/g < h\;.$}
		 \end{array}
	 \right. \]
For $g$ less than some critical value $g_c$, activity $K^*$
increases with noise as the system settles onto the stable fixed
point $K^* = 1 + \sigma^2$, whereas for $g > g_c$ chaos sets in. It
was found that for small noise asymptotically
$g_c = 1 - \sigma^2 \ln\;\sigma^2$ and for large noise
$g_c = \sqrt{\pi/2}\;\sigma$. Fig.~1 is the plot of
$g_c$ verses $\sigma$ (phase diagram),
which clearly shows the chaotic and the regular regimes.

MSS concluded that for higher dimensional systems of this
type where chaos occurs essentially due to the randomized interactions
amongst constituent nonchaotic elements, noise acts to impair
information flow between these elements and thereby suppresses chaos.

\subsection{Spatiotemporal stochastic resonance in excitable media}

Let us now consider the observation of stochastic resonance phenomenon
in a spatially extended pattern forming system, as reported by
Jung and Mayer--Kress (JMK) \cite{JM}.
Stochastic resonance, as the name suggests, is a phenomenon 
in which at a given noise--strength certain temporal patterns of
the system get enhanced drastically owing to an increase
in the system's sensitivity \cite{BSV,MnW,FL}.
This behaviour has a characteristic bell--shaped curve 
when the output power (of fourier spectrum) at the corresponding
frequency is plotted against noise--strength, with the peak at 
the `resonating' noise value.
In order to observe this type of phenomenon in an extended system
JMK considered a two--dimensional equidistant square array (with lattice
constant $a$) of $N \times N$ noisy threshold devices
$d(i,j), i,j = 1,\cdots,N$ with the
following properties: if the input $x_t(i,j)$ to the devices is below
some threshold $b$,
output $y_t(i,j) = 0$; if $x_t(i,j) > b$ the device generates a spiky
output of the form $y_t(i,j) = I_0 \Theta(\dot{x}(i,j))\;\delta(x_t(i,j) - b)$
and then goes into temporary hibernation for a refractory period
$\Delta t_r$. The dynamics can be described by a linear Langevin equation
for inputs $x_t(i,j)$ as
\begin{eqnarray}
\dot{x}(i,j) = -\gamma x(i,j) + \sqrt{\gamma \sigma}\;\eta_t(i,j)\;,
							\label{stsr1}
\end{eqnarray}
where $\gamma$ is a leakage constant which accounts for the thermal leakage
due to interaction with the surroundings, $\eta_t(i,j)$ is a delta--correlated
noise and $\sigma = \left< x_t(i,j)^2 \right>$. Eq.~(\ref{stsr1})
on integration over $\Delta t$ gives rise to the discrete time dynamics
\begin{eqnarray}
x_{t+\Delta t}(i,j) = x_t(i,j)\;{\rm exp}(-\gamma \Delta t) + G_t(i,j)\;,
							\label{stsr2}
\end{eqnarray}
where $G_t(i,j)$ is a Gaussian random number with variance 
$\sigma_G = \sigma[1 - e^{-2 \gamma \Delta t}]$.

Threshold devices are pulse--coupled, {\it i.e.}, when an element $d(k,l)$
fires, it communicates with its surrounding elements $d(i,j)$ at a distance
$r_{ij,kl}$ apart by contributing an amount
$K\;{\rm exp}(- \beta r^2_{ij,kl}/a^2)$ to their inputs $x_t(i,j)$,
where $K \rightarrow K/b$ is in dimensionless unit and $\beta$ is a
dimensionless quantity describing the spatial coupling range.
For large coupling $(K \geq {\rm exp}(\beta))$, these firing elements give
rise to excitatory waves spreading through the array, such as spiral
waves, target waves (single, nonrepetitive wavefronts) etc. The selection
of the wave form depends on the geometry of initial conditions.
There is a threshold
$K_c$ such that for $K < K_c$ excitatory waves do not necessarily start.

However, in the presence of noise the spreading of excitatory waves,
such as spiral waves, is observed
in subthreshold regime $(K < K_c)$. Any such noise--sustained wave
disappears soon after the noise is turned off. With increasing noise
the spiral wave evolves with a larger curvature, $i.e.$, with an
enhanced coherence in the array. If the noise is increased further,
the spiral breaks up and there is a loss of coherence.

In order to show that the phenomenon observed above is essentially
spatiotemporal stochastic resonance, JMK have taken a solitary 
initial wave and defined a quantity $\mu$, time averaged number of
excess events, as the difference of number of firing elements
along the driving wavefront and average number of firing events along a
row of the array, not affected by the driving. The plot of $\mu$ versus
$\sigma$, the variance of noise, is shown in figure 2. The plot exhibits
the characteristic bell--shaped curve mentioned earlier.
Synchronization of the spatiotemporal
firing pattern to an external driving, thereby giving rise to the
resonating peak, is a simple generalization of the stochastic resonance
to this extended system.

We shall now move on to the final part of the paper,
effect of noise on coupled map lattices (CML). As the model itself has
a plethora of interesting dynamical features, we begin with an
introduction on CML.

\section{Coupled map lattice}

Coupled map lattices (CML) are spatially extended
dynamical systems with discrete space (lattice), discrete time
and continuous state evolution. General evolutionary
dynamics of a CML may be expressed as
\begin{eqnarray}
x_{t+1}(i) = F(x_t(i))
  + \sum_{j \neq i} \varepsilon_{ij} G(x_t(j),x_t(i))\;, \label{cml1}
\end{eqnarray}
where $F$ and $G$ are nonlinear maps
and the state variables $x_i$ varies continuously in the phase space
of the map and $i$ is the space index. CMLs are in general constructed with the symmetry properties of
spatial translational and rotational invariance, and so the coupling
term $\varepsilon_{ij}$ is taken to be a uniform scalar ranging from
the nearest 
neighbour coupling to global coupling.

CMLs have been used to model many physical phenomena. These include
pattern formation, chemical waves, excitable media, nucleation,
crystal growth, charge density waves, population dynamics, fluid
dynamics etc. [42--55]. Chemical systems exhibit
very rich spatiotemporal structures and patterns. Standard description
of these systems is in terms of the reaction--diffusion equations that
incorporate combined effects of local nonlinear reaction dynamics and
global diffusion of chemical species due to the concentration gradients
in the system. Many of the averaged and even some detailed features
of chemical patterns can be captured by a simple CML description
of the dynamics. Another area of interest is the phase ordering dynamics.
Phase separation occurs due to competition between different states in
the system, such as domain growth in ferromagnetic and chemical systems.
Some of the salient features of phase ordering dynamics are also 
observed in CML model. CML may also be applied in pattern dynamics with
an excitable state and a relaxation from it. Such processes are observed
in reaction--diffusion in excitable media and also in biological problems
such as heart rhythm and electrical activities in neural tissues. In
fluid dynamics, simulations using CML can show formation of convective
patterns, vortices, sinks, sources etc. Here CML may offer a
computationally economical way of simulating real behaviour. One of the
major interests in studying CML is in the context of understanding the
spatial and
temporal structure formation, specially
in fluid dynamics. These structures appear in spite of the fact that
the underlying evolution of the system is spatiotemporally chaotic.

One of the most extensively studied types of one dimensional CML is that with
nearest--neighbour coupling and having the following form:
\begin{eqnarray}
x_{t+1}(i) = F(x_t(i)) + \frac{\varepsilon}{2}\;{\big [}G(x_t(i-1))
		+ G(x_t(i+1)) - 2 G(x_t(i)){\big ]}\;. \label{cml2}
\end{eqnarray}
We shall
concentrate on the form $G(x) = F(x)$, which reduces Eq.~(\ref{cml2})
to
\begin{eqnarray}
x_{t+1}(i) = (1 - \varepsilon)F(x_t(i)) + 
   \frac{\varepsilon}{2}{\big [}F(x_t(i-1)) + F(x_t(i+1)){\big ]}\;.
							\label{cml3}
\end{eqnarray}
This particular model is known as {\it future diffusive} CML since
the diffusively coupled entities are one timestep evolved values.
Studies of this model alongwith the local
dynamics as logistic map $F(x) = \mu x (1 - x)$ have revealed that
they can exhibit a wide range of spatiotemporal complexity. It has
been observed that the temporal period doubling behaviour of the map
can induce spatial domain structures separated by kinks and antikinks.
A pattern selection regime is observed where patterns of certain 
characteristic lengths are selected. Phenomena such as spatiotemporal
quasiperiodicity, soliton turbulence, spatiotemporal intermittency,
wavelength doubling bifurcations, synchronization etc. for this
and other maps
(like circle map $\theta_{t+1} = \theta_t + \Omega - \frac{K}{2 \pi}
{\rm sin} (2 \pi \theta_t)$, $\Omega$ -- angular frequency,
$K$ -- nonlinearity parameter)
have been identified [56--64].

We now discuss the effect of noise on the evolution of CML. In the next
subsection we report
a novel phenomenon that we have observed in this context \cite{RA}.

\subsection{Noise and `stochastic coherence' in CML}

We take the system of Eq.~(\ref{cml3}) with an
additional noise term as follows:
\begin{eqnarray}
x_{t+1}(i) = (1 - \varepsilon)F(x_t(i)) + \frac{\varepsilon}{2}
   {\big [}F(x_t(i-1)) + F(x_t(i+1)){\big ]} + \eta_t\;,
							\label{sc1}
\end{eqnarray}
where $\eta_t$ is the familiar delta--correlated noise.
Logistic function $F(x) = \mu x (1 - x)$ is used as local dynamics.
For noise $\eta_t$ we have used 
a uniformly distributed random number bounded between $-W$ and $+W$,
with the constraint that $0 < x_{t+1}(i) < 1$;
we call $W$ the noise--strength parameter. 
Values of $\mu$, $\varepsilon$, and $L$ (size of the lattice) are
chosen so that the resultant dynamics of the system is chaotic.

Now, we define a {\it structure} as a region of space such that the difference
in the values of the variables of neighbouring sites within this region is
less than a predefined small positive number say $\delta$, $i.e.$,
$\vert x_t(i) - x_t(i\pm1) \vert \leq \delta$.
We call $\delta$ the
structure parameter. We look for such patterns, or
structures, to appear in the course of
evolution of the model given by Eq.~(\ref{sc1}).

Figure 3 shows a plot (on log--log scale) of the distribution of the 
number $n(l)$ vs. length $l$ 
of the structures for different values of $W$, with $\mu = 4.0$,
$\epsilon = 0.6$, 
$\delta = 0.0001$ and $L = 1000$, and open--boundary conditions are used. 
Power--law nature of the decay of $n(l)$ is 
clearly evident, which has a form
\begin{eqnarray}
n(l) \propto l^{-\alpha}\;, \label{sc2}
\end{eqnarray}
where $\alpha$ is the power--law exponent.
This indicates that the system does not have any intrinsic
length scale.
It may be noted that in the absence of noise ($W = 0.0$) the decay is  
manifestly exponential \cite{JA}.
Exponent $\alpha$ is seen to depend on the
noise--strength $W$, with a minimum for $W$ around 0.6. 
We define average length $\bar{l}$ of a structure as 
$\bar{l} = \sum{l\,n(l)} / \sum{n(l)}$. In Fig.~4 we plot the variation of
$\bar{l}$ with $W$ for values of parameters as in Fig.~3.
The plot exhibits a {\it bell--shaped} nature within a fairly 
narrow range of $W$ around value 0.6 (one may note the surprising
similarity between figures 4 and 2, though they refer to completely
different phenomena). It may be noted that the minimum
of $\alpha$ also occurs for $W$ quite close to this value, as expected. 

We call the phenomenon observed above {\it stochastic coherence}. This is
similar to
stochastic resonance which shows a bell--shaped behaviour of 
temporal response as a function of the noise--strength, as mentioned before. 
However one 
may note that our system does not have any intrinsic length--scale, whereas
in stochastic resonance noise resonates with a given time--scale;
hence our use of the word coherence rather than resonance.
In stochastic resonance
noise transfers energy to the system at a characteristic frequency,
whereas in stochastic coherence noise induces coherence to the system. 

To study the evolutionary aspects of these
structures we obtained distribution of number $n(\tau)$ of structures vs. 
their lifetimes $\tau$ for different $W$.
$n(\tau)$ is observed to decrease with $\tau$ with a 
stretched exponential type of decay having a form 
\begin{eqnarray}
n(\tau) \propto \exp\big(-({\rm const.})\tau^\zeta\big)\;, \label{sc4}
\end{eqnarray}
where $\zeta$ depends on $W$. 
We define average lifetime
$\bar{\tau}$ of a structure as
$\bar{\tau} = \sum{\tau\,n(\tau)} / \sum{n(\tau)}$. In Fig.~5 we plot
$\bar{\tau}$
vs. $W$ for parameter values as in Fig.~3. The graph shows a
bell--shaped
feature with maximum for 
$W$ around $0.6$.

In order to ascertain the chaotic nature of the system
evolution we have calculated
the lyapunov exponent spectra for our system. We find a number of lyapunov
exponents
to be positive, implying that the underlying evolution is chaotic.
Maximum lyapunov 
exponent $\lambda$ shows a minimum around noise--strength $0.6$
The fact that we have observed other extrema for similar $W$ may
make it appear that probably the origin of these behaviours
lies with reduction of $\lambda$ due to noise \cite{MT} (though
it should be noted that the reduction is not monotonic).
To explore this possibility
we have studied the variation of $\lambda$ with coupling parameter
$\varepsilon$. 
We found that
$\lambda$ remains fairly constant for $0.2 \leq \varepsilon
\leq 0.8$
for the entire range of $W$. On the other hand, a plot of variation of average
length $\bar{l}$ with $\varepsilon$ for fixed $W$ shows a monotonically
increasing behaviour, quite contraty to what is expected from 
$\lambda$. This implies that the lyapunov exponent alone cannot
be used for proper characterization of spatio--temporal features of the 
system (unlike the synchronization phenomenon discussed earlier, where
$\lambda$ is sufficient to fully characterize the behaviour).

To conclude this section, we have reported a new phenomenon observed
in a chaotically  
evolving one--dimensional CML driven by an identical noise, which we termed
stochastic coherence. It is observed that there is a phenomenal increase 
in the abundance of coherent structures of all scales due to noise.
Distribution of these structures shows a power--law decay with length of the 
structure.
Average length
as well as average lifetime of these structures exhibit characteristic maxima
at certain noise--strength.

\section{Summary and conclusion}

We have reviewed the work that has been carried out on the effect of
noise on evolutionary dynamics of chaotic systems.
In the case of uncoupled systems,
we have encountered synchronization phenomenon for different phase space 
trajectories driven by identical noise above a certain strength.
We have also seen nonstatistical behaviour for an ensemble of
chaotic trajectories driven by an identical sequence of parametric noise.
For coupled systems, noise above a given strength has been observed to suppress
chaos in a random neural network with infinite range interactions.
Spatiotemporal stochastic resonance, first case of stochastic resonance
phenomenon observed in systems with spatial extension, has been
reported for a two dimensional square array of firing devices. In the context
of CML, we have observed an interesting phenomenon, stochastic coherence,
when the entire lattice is driven by identical noise.

As noted earlier in the text, understanding formation and 
evolution of patterns and structures in the spatially extended systems is
still incomplete. In real physical situations, noise almost
ubiquitously influences the system behaviour. So while studying the ordering
phenomena, 
incorporating noise as a part of the evolving `supersystem' and 
studying mutual interaction of the two subsystems
offers a more practical
way of looking into the problem. The inherent nonlinearity in the system,
which on one hand gives rise to chaos and on the other hand 
ordered behaviour,
is surely again playing its role in enhancing order under the
influence of noise; how so is still not quite very clear.
We have 
encountered two kinds of noise induced ordering phenomena: global
ordering (such as synchronization) and local ordering (spatiotemporal
stochastic resonance and stochastic coherence). Stochastic coherence,
in particular,
may turn out to be a promising way towards achieving any degree of
understanding of the structure formation in extended systems, although
a lot of work remains to be done. While it is 
definitely not claimed that all questions will be answered with this
approach, we certainly hope to bridge a few gaps in the existing knowledge.

\vskip 0.3in

\noindent
{\Large \bf Acknowledgments}

\vskip 0.2in

\noindent
One of the authors (M.R.) acknowledges University Grants Commission (India)
and the
other (R.E.A.) acknowledges Department of Science and Technology (India) for 
financial assistance.

\newpage

\begin{center}

{\bf Figure Captions}

\end{center}

\begin{itemize}

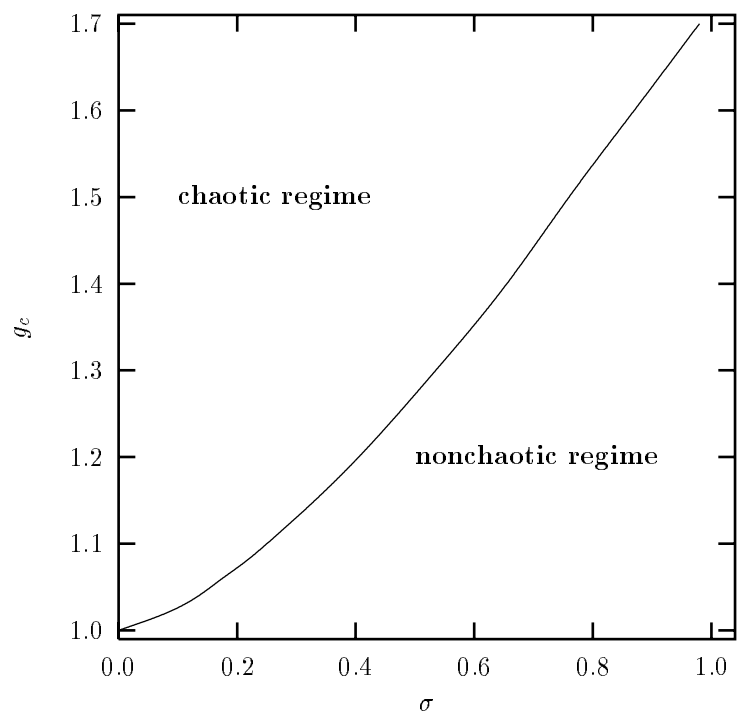
\item[Fig.~1.] Plot of phase diagram ($g_c$ versus $\sigma$)
obtained by Molgedey $et. al.$ \cite{MSS} for an
infinite range neural network with external white noise. For a given
gain $g$ chaos is suppressed for sufficiently strong noise $\sigma$.

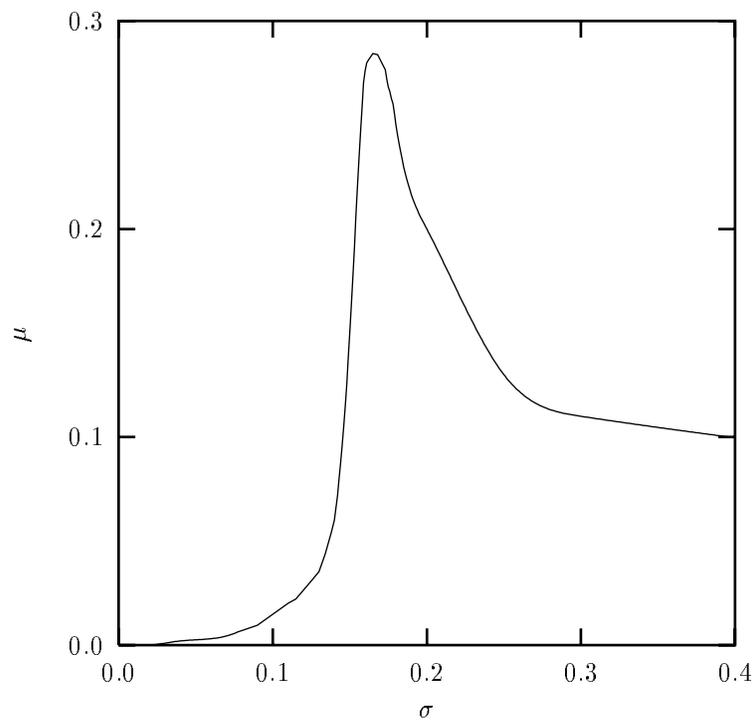
\item[Fig.~2.] Plot of the graph of time averaged
number of excess events $\mu$ with noise
variance $\sigma$, for a typical lattice
size $200 \times 200$ and with a given set of parameter values \cite{JM}.

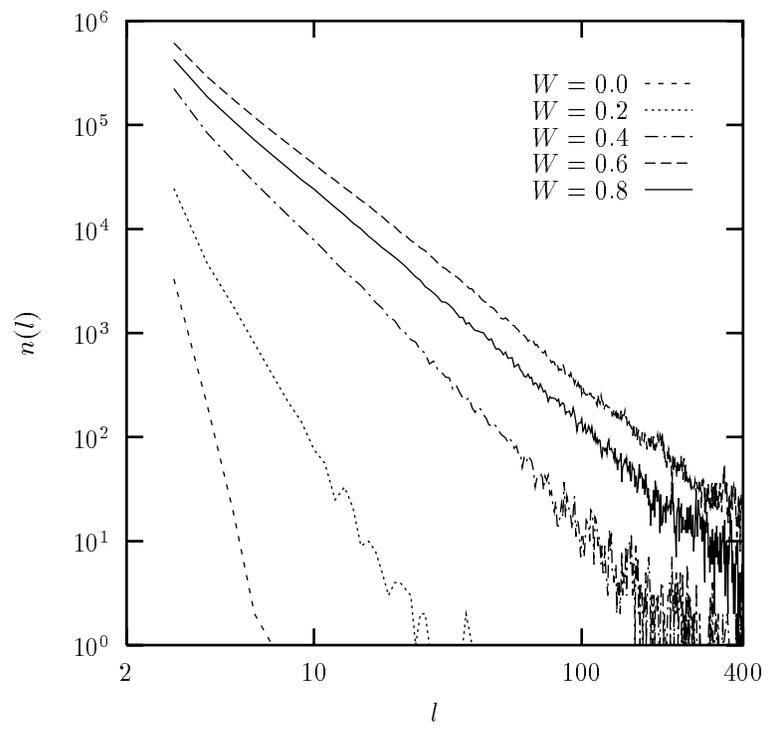
\item[Fig.~3.] Plot of variation of number $n(l)$ of structures 
with length $l$ for 
a lattice with size $L = 1000$, for different values of noise--strength $W$ as 
indicated.
Parameters chosen are
$\varepsilon = 0.6$, $\delta = 0.0001$, and
$\mu = 4.0$. Open--boundary conditions are used. Data are obtained for 50000 
iterates each for 10 initial condition.

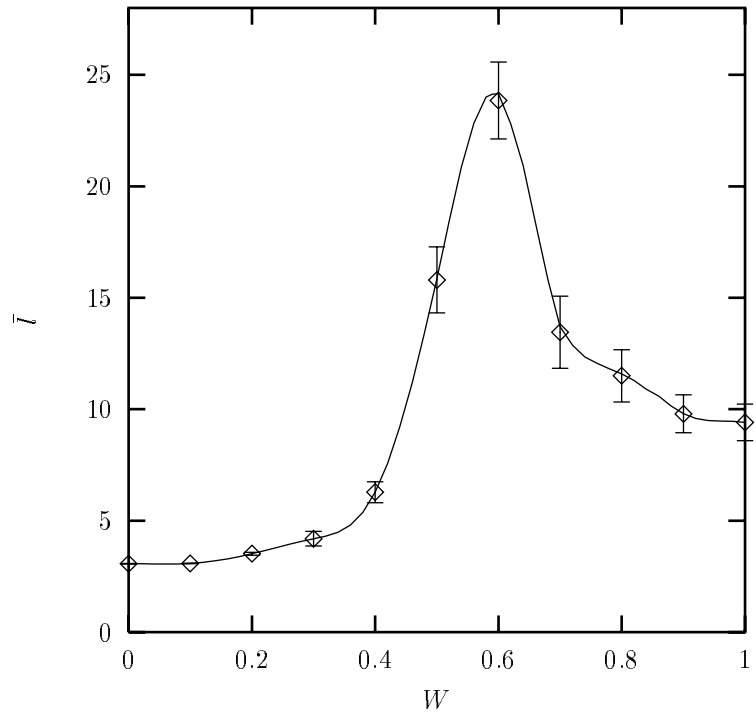
\item[Fig.~4.] Plot of variation of average length $\bar{l}$ of structure with 
noise--strength $W$, with parameters as stated in Fig.~3.

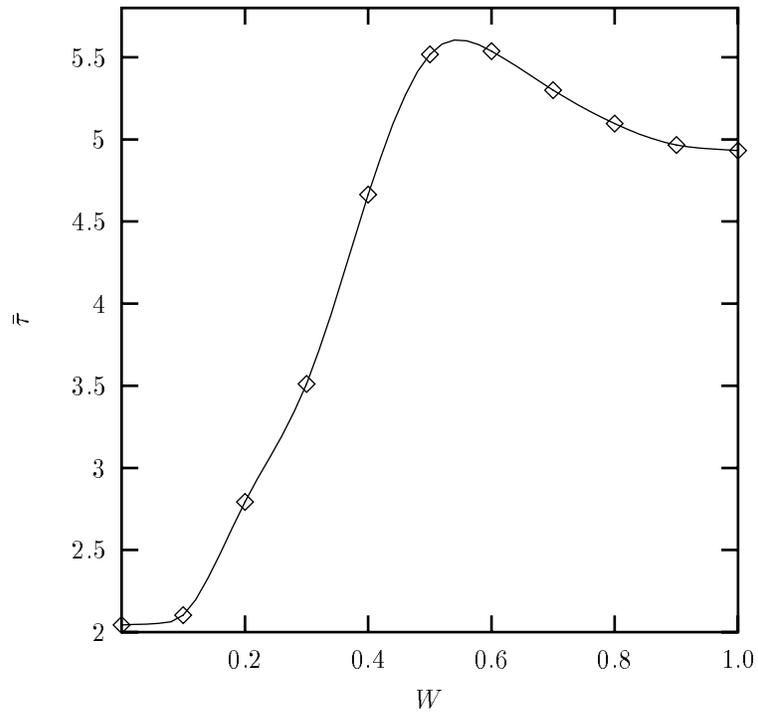
\item[Fig.~5.] Variation of average lifetime $\bar{\tau}$ of structures with 
noise--strength
$W$ shown for parameters as stated in Fig.~3.

\end{itemize}

\end{document}